\documentclass[sigconf,authorversion,nonacm]{acmart}
\AtBeginDocument{%
  \providecommand\BibTeX{{%
    Bib\TeX}}}

\usepackage[caption=false]{subfig}
\usepackage{amsmath,amsfonts}
\DeclareMathOperator*{\argmax}{arg\,max}
\DeclareMathOperator*{\argmin}{arg\,min}
\usepackage[ruled, vlined, linesnumbered]{algorithm2e}
\usepackage{graphicx}
\usepackage{textcomp}
\usepackage{xcolor}
\def\BibTeX{{\rm B\kern-.05em{\sc i\kern-.025em b}\kern-.08em
    T\kern-.1667em\lower.7ex\hbox{E}\kern-.125emX}}
    
\makeatletter
\def\@copyrightspace{\relax}
\makeatother
\begin{document}

\title{Accelerating Deep Neural Network guided MCTS using \\ Adaptive Parallelism}


\author{Yuan Meng}
\authornote{Both authors contributed equally to this research.}
\email{ymeng643@usc.edu}
\orcid{0000-0001-6468-8623}
\affiliation{%
  \institution{University of southern California}
  \country{USA}
}

\author{Qian Wang}
\authornotemark[1]
\email{pwang649@usc.edu}
\orcid{0009-0003-6157-2459}
\affiliation{%
  \institution{University of southern California}
  \country{USA}
}

\author{Tianxin Zu}
\email{zut@usc.edu}
\affiliation{%
  \institution{University of southern California}
  \country{USA}
}

\author{Viktor Prasanna}
\email{prasanna@usc.edu}
\affiliation{%
  \institution{University of southern California}
  \country{USA}
}


\begin{abstract}
Deep Neural Network guided Monte-Carlo Tree Search (DNN-MCTS) is a powerful class of AI algorithms.
In DNN-MCTS, a Deep Neural Network model is trained collaboratively with a dynamic Monte-Carlo search tree to guide the agent towards actions that yields the highest returns. While the DNN operations are highly parallelizable, the search tree operations involved in MCTS are sequential and often become the system bottleneck.
Existing MCTS parallel schemes on shared-memory multi-core CPU platforms either exploit data parallelism but sacrifice memory access latency, or take advantage of local cache for low-latency memory accesses but constrain the tree search to a single thread. In this work, we analyze the tradeoff of these parallel schemes and develop performance models for both parallel schemes based on the application and hardware parameters. We propose a novel implementation that addresses the tradeoff by adaptively choosing the optimal parallel scheme for the MCTS component on the CPU. Furthermore, we propose an efficient method for searching the optimal communication batch size as the MCTS component on the CPU interfaces with DNN operations offloaded to an accelerator (GPU). Using a representative DNN-MCTS algorithm - Alphazero on board game benchmarks, we show that the parallel framework is able to adaptively generate the best-performing parallel implementation, leading to a range of $1.5\times - 3\times$ speedup compared with the baseline methods on CPU and CPU-GPU platforms.
\end{abstract}



\maketitle

\section{Introduction}
Deep Neural Network guided Monte Carlo Tree Search (DNN-MCTS)  methods have shown massive potential in modern AI benchmarks. For example, DNN-MCTS is the core in state-of-the-art algorithms, including Alphazero \cite{alphazero} in gaming, AlphaX \cite{wang2020neural} in neural architecture search, CAPR \cite{chen2019context} in recommendation systems, etc. In traditional MCTS, an agent ``looks ahead'' the future scenarios by constructing and traversing a partial search tree. In the search tree, nodes correspond to states, and edges represent actions performed by the agent. 
The key objective of an MCTS algorithm is guiding the partial tree traversal so that the agent can focus on more important nodes leading towards high rewards. 
To evaluate the importance of nodes to be included in the partial tree, Monte-Carlo rollouts \cite{coulom2006efficient} are adopted in traditional MCTS, where a possible outcome is sampled from the state by simulating from the state using an application-specific environment simulator. DNN-MCTS improves upon the
traditional MCTS by eliminating such Monte-Carlo rollouts.
Instead of simulations, in DNN-MCTS \cite{alphazero}, a node is evaluated using a Deep Neural Network (DNN) trained on data sets collected online through tree-based search. This not only enables high algorithm performance without prior human knowledge but also replaces sequential, application-specific simulation steps with dense tensor operations, which leads to ample opportunities for parallelization and hardware acceleration.

Training the DNN using MCTS is an extremely time-consuming process. For example, a DNN-MCTS algorithm on the Go game benchmark, AlphaGo Zero, was trained for 21 days \cite{silver2017mastering}. Thus, enabling faster DNN-MCTS training is an important problem.
In DNN-MCTS, the DNN is collaboratively trained with the tree. Specifically, data points collected during the MCTS tree-based search (with simulated final outcomes as ground truth) are used for updating the DNN parameters; the value approximations returned by DNN inferences are used for updating the tree nodes in the Monte Carlo search tree during the tree-based search.
In our initial experiments, the tree-based search accounts for more than 85\% of the total runtime in an iteration of serial DNN-MCTS.
A popular parallel algorithm for accelerating the tree-based search process is tree-parallel DNN-MCTS, it is widely adopted in many DNN-MCTS implementations such as AlphaZero \cite{silver2017mastering} and AlphaX \cite{wang2019alphax}. 

In the tree-based search process of tree-parallel DNN-MCTS, even though the independent  DNN inferences from multiple nodes can be executed in a data-parallel manner, it is challenging to obtain linearly-scalable speedups wrt the number of processes allocated to parallel workers. This is because multiple processes sharing the same tree either require frequent synchronizations or are completely serialized to preserve the most up-to-date node parameters for accurate node selection.

In this paper, we propose an adaptive-parallel methodology for tree-parallel DNN-MCTS based on an analysis of tradeoffs between two parallel implementations (local-tree and shared-tree). We target the tree-based search process of DNN-MCTS, which involves in-tree operations and DNN inferences. We optimize the MCTS in-tree operations on a shared-memory multi-core CPU architecture. Our implementation support GPU accelerated DNN inferences. Our contributions are:
\begin{itemize}
    
    \item  We perform the tradeoff analysis between the two implementations (shared-tree and local-tree methods) and propose an acceleration methodology of adaptively selecting the implementation given an arbitrary DNN-MCTS algorithm targeting a multi-core CPU.
    \item We implement both local-tree and shared-tree parallel DNN-MCTS as a single program template that allows compile-time adaptive selection of parallel implementations; the program template allows interfacing with existing high-level libraries for simulating various benchmarks, and supports offloading the DNN computations to accelerators.
    \item We propose a design configuration workflow that decides the optimal parallel method at compile time. This is achieved using high-level performance models for two tree-parallel DNN-MCTS implementations based on algorithm hyper-parameters (e.g., tree fanout, tree depth), hardware specifications (e.g., number of threads, DDR bandwidth and latency), and design-time profiling. 
    \item We utilize an efficient search method that determines the best DNN-request-processing batch size in the design configuration workflow to fine-tune the DNN-MCTS performance on a CPU-GPU platform. 
    This is achieved by overlapping DNN request transfers with in-tree operations and minimizing the GPU wait time.
    \item We successfully validated the proposed adaptive parallel methodology by running the Gomoku board-game benchmark and achieved up to $3\times$ speedup than the baselines using either parallel implementation alone.
\end{itemize}
\section{Background}
\subsection{DNN-MCTS}
The complete DNN-MCTS training pipeline is an iterative process composed of two stages: tree-based search and DNN training. The tree-based search stage is guided by the DNN inference results on a tree, and generates the datasets used for DNN training.
The DNN takes the current state $s$ as the input, and outputs a value estimation of $s$ and a policy (i.e., the probabilities of taking each available action from $s$).
Each node in the tree represents a certain environment state. Each edge represents the action that transits from one state to another, and tracks the visit counts and application-specific values associated with the action. For example, in AlphaZero \cite{alphazero}, each edge maintains $Q(s,a)$ - the expected reward (i.e. the Q value) for taking action $a$ from state $s$;
$N(s,a)$ - the number of times action $a$ is taken from state $s$ in all the iterations in a search stage; $P(s,\cdot)$ - the policy returned by the DNN, which is the probability of taking each action from the state $s$. 

In the tree-based search stage, each iteration of the tree-based search is composed of the following operations:
\begin{enumerate}
    \item \textbf{Node Selection:} The search starts from the current state (root node of the tree) and traverses down the tree. At every node traversed $s$, the next edge is selected according to the statistics stored in the search tree as follows:
    \begin{multline}
    \label{eq:uct}
    a=argmx(U(s, a)), \text{where the UCT score}\\
    U(s, a)=Q(s, a)+c\cdot P(s, a) \cdot \frac{\sqrt{\Sigma_b N(s, b)}}{1+N(s, a)}
    \end{multline}
    This leads the agents towards states with high reward values (exploitation), high policy-action probability, and low visit counts (exploration). $c$ is a pre-set constant controlling the tradeoff between exploitation and exploration. 
    \item \textbf{Node Expansion \& Evaluation:}
    When the tree traversal encounters an edge that was never visited before, the search process adds a new successor node $s'$, and initializes $Q(s',a), N(s',a)$ to 0 for all its adjacent edges $a$. Accordingly, $P(s',\cdot)$ is derived from the DNN inference which takes the new node $s'$ as input; the DNN also outputs the estimated reward value  $v(s',\cdot)$.
    \item \textbf{Backup:} To synchronize the tree with the most recent node evaluation, $v(s',\cdot)$ is propagated from the new leaf node back to the root. At each tree level, the visit counts $N$ is incremented, and the state value $Q$ is accumulated using $v$.
\end{enumerate}
After a fixed amount of iterations, the best move is picked at the root node (i.e., the current state $s_t$) based on Equation \ref{eq:uct}. This generates a training datapoint $(s_t, \vec{\pi}_t,r)$, where $\vec{\pi}_t$ is the action statistics at the root, and $r$ is the reward recorded at terminal states. These training data points are later consumed by the DNN training stage.

In the DNN training stage, the DNN performs a stochastic gradient descent (SGD, \cite{robbins1951stochastic}) using the data points generated in the tree-based search state. For example, In AlphaZero \cite{alphazero}, it updates the DNN parameters $\theta$ to minimizes the loss:
\begin{equation}
\label{eq:loss}
    l=\sum_t\left(v_\theta\left(s_t\right)-r\right)^2-\vec{\pi}_t \cdot \log \left(\vec{p}_\theta\left(s_t\right)\right)
\end{equation}
where $v_\theta$ and $p_\theta$ are the value head and policy head of the DNN output.

In our initial profiling of the sequential DNN-MCTS on Gomoku benchmarks \cite{tan2009adaptive}, the tree-based search stage account for more than 85\% of the complete training process. Therefore, there is a critical need for parallelizing both the MCTS and DNN inference processes in the tree-based search stage. \textit{ Our work focus on the (variations of) Tree Parallelization \cite{treeP, wu-uct}}. This is recently the most popular MCTS parallelization technique used in existing DNN-MCTS implementations such as AlphaZero \cite{alphazero}.
In Tree-Parallel MCTS, after a worker traverses a certain node (path) during Node Selection, a virtual loss \textit{VL} is subtracted from $U$ of the traversed edges to lower their weights, thus encouraging other workers to take different paths. It also creates dependencies between workers during the Node Selection. \textit{VL} is recovered later in the BackUp phase. Note that \textit{VL} can either be a pre-defined constant value \cite{treeP}, or a number tracking visit counts of child nodes \cite{wu-uct}. 

In this work, we view the tree-based search stage as a composition of in-tree operations and DNN inference. The in-tree operations are all the operations that access the tree in Node Selection, Node Expansion, and BackUp phases, and the DNN inference refers to Node Evaluation. Note that the target platform for in-tree operations is a multi-core CPU, and DNN inference may be executed on the CPU or offloaded to an accelerator.

\subsection{Related Work}
Other than tree-parallel MCTS targeted in this work, multiple other parallel algorithms have been developed for high-throughput MCTS and DNN-MCTS.
Leaf-parallel MCTS \cite{leafP} uses a single tree and creates multiple parallel node simulations at the same leaf node, but it wastes parallelism due to the lack of diverse evaluation coverage on different selected paths,
which leads to algorithm performance degrades \cite{kato2010parallel}. 
Root-parallel MCTS \cite{rootP} creates multiple trees at different workers and aggregates their statistics periodically, but still lets multiple workers visit repetitive states. 
The Speculated DNN-MCTS \cite{kim2021specmcts} comply with the sequential in-tree operations, and uses a speculative model in addition to the main model for faster node evaluation. This preserves the decision-making quality of the sequential MCTS but introduces additional computations.

The original tree-Parallel MCTS \cite{treeP} uses multiple workers to share and modify the same tree, and uses mutex to avoid race conditions. However, the synchronization overhead can dominate the memory-bound in-tree operations, making the achievable speedups sub-optimal.
\cite{mirsoleimani2018lock} attempts to address this by developing a lock-free tree-parallel method, but the agents trained cannot win against root-parallel MCTS on hex game benchmarks without careful tuning of hyper-parameters.
WU-UCT \cite{wu-uct} puts multiple workers on the same thread and executes them in a centralized manner using a local tree, while parallelizing the node evaluations (simulations). This avoids overheads from frequent thread-synchronizations, but the speedup does not linearly scale up wrt allocated parallel resource when the sequential workers become the bottleneck \cite{meng2022accelerating,meng2023framework}.
Overall, there are different tradeoffs wrt the execution speed of the best-performing agents. Therefore, we are motivated to combine the different advantages of a tree-Parallel MCTS with shared tree \cite{treeP} and local tree \cite{wu-uct}, and dynamically select between them to suit different scenarios.


\begin{figure}[h]
    \centering
    \subfloat[Shared-tree on multi-core system]{%
    \includegraphics[clip,width=0.5\columnwidth]{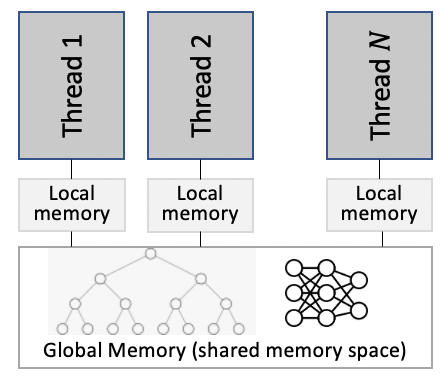}%
    }
    
    \subfloat[Execution timeline of the shared-tree method]{%
    \includegraphics[clip,width=\columnwidth]{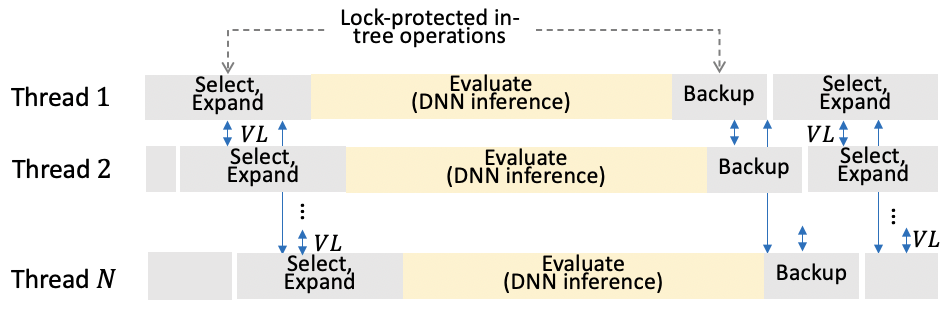}%
    }
    \caption{Shared-tree method}
    \label{fig:decen_timeline}
\end{figure}
\section{Parallelization Schemes and Implementation}
\subsection{Parallelization Schemes}
Assume that we allocate $N$ workers sharing the tree during the tree-based search. We consider two methods to implement tree-parallel MCTS on multi-core CPUs. These methods are characterized by their usage of a local tree and a shared tree, respectively:

\subsubsection{Shared Tree}
The shared-tree method uses $N$ threads in total - it assigns each worker an individual thread.  Each thread is responsible for its own assigned worker's in-tree operations and DNN inference. The tree is stored in a shared memory (typically DDR memory of the CPU), and nodes in the tree are distributed to parallel workers as they access the tree. The shared-tree method on a multi-core system is shown in Figure \ref{fig:decen_timeline}-(a). The in-tree operations by each work are protected with locks so that only one worker can access a certain node at a time. The operation execution timeline of the shared-tree method is shown in Figure \ref{fig:decen_timeline}-(b). All workers start at a common root node, and the virtual loss applied to the root children needs to be updated for all workers accessing it. So, the time interval between consecutive workers involves the overhead for communicating the root-level information through share memory space (i.e., DDR), creating latency offsets between workers. The main advantage of the shared-tree method is that in-tree operations are parallelized. The disadvantage is that the more compute-intensive Node Evaluation process cannot fully utilize the compute power provided by the parallel threads, since they need to wait for the completion of in-tree operations by all workers, and these in-tree operations are bounded by memory access latencies.

\subsubsection{Local Tree}

\begin{figure}[h]
    \centering
    \subfloat[Local-tree on multi-core system]{%
    \includegraphics[clip,width=0.7\columnwidth]{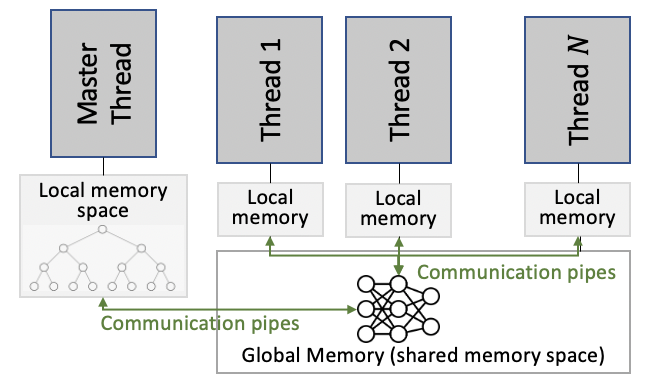}%
    }
    
    \subfloat[Execution timeline of the local-tree method]{%
    \includegraphics[clip,width=\columnwidth]{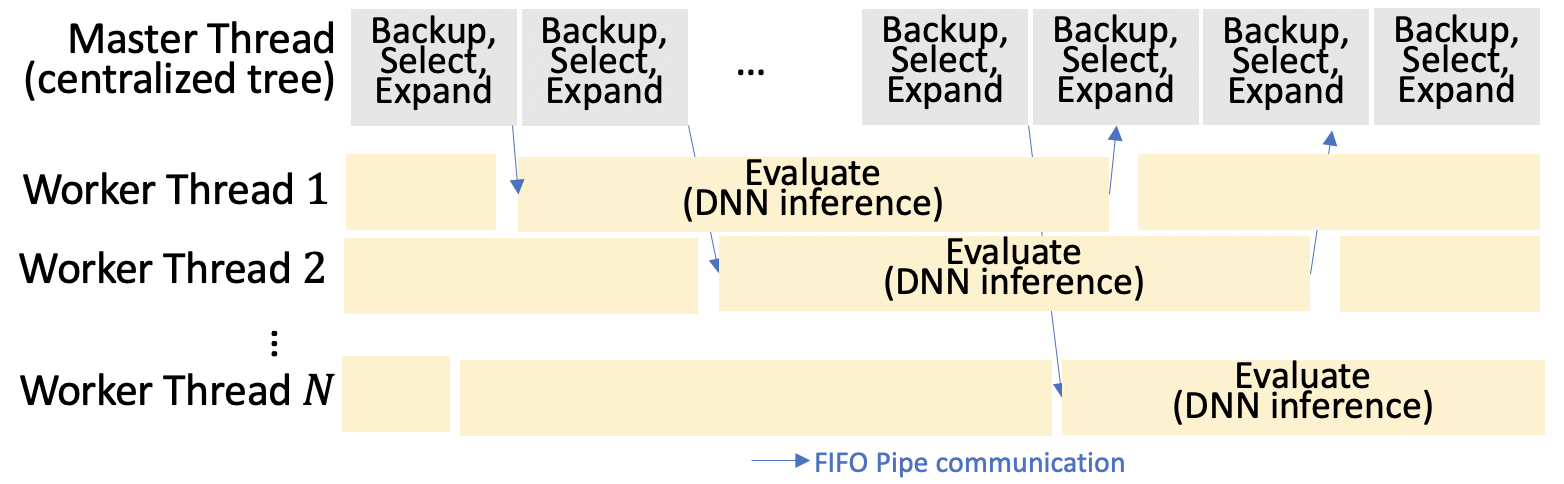}%
    }
    \caption{Local-tree method}
    \label{fig:local_method}
\end{figure}

\begin{algorithm}
\DontPrintSemicolon
\caption{Adaptive Parallel DNN-MCTS}
\SetKwFunction{FMain}{main}
  \SetKwProg{Fn}{Function}{:}{}
  \Fn{\FMain{$flag\_local$}}{
  \For{\_\_ in training\_episodes} {
        Initialize $environment$\;
        Initialize $dataset$\;
        \While{not environment.terminal} {
            \If{$flag\_local$}{
                $ap \gets$ \texttt{get\_action\_prior\_l} ($environment$)\;
            }
            \Else{
                $ap \gets$ \texttt{get\_action\_prior\_s} ($environment$)\;
            }
            take action $\argmax$ $ (ap)$\;
            $reward \gets$ update ($environment.state$)\;
            $dataset$.append ($environment.state$, $ap$, $reward$)
        }
        \For{\_\_ in SGD\_iterations} {
            batch $\gets$ sample($dataset$)\;
            SGD\_Train(batch)\;
        }
    }
  }
\label{alg:1}
\end{algorithm}
The local-tree method uses $N+1$ threads in total - it uses a centralized master thread to manage the complete tree, and it allocates $N$ threads to execute the Node Evaluations for $N$ workers (each thread is solely dedicated to the DNN inferences). The complete tree is stored in the local memory of the master thread (e.g., cache memory). The master thread also manages a worker-thread pool where the master thread communicates with each worker thread through a FIFO (first-in-first-out) communication pipe. The local-tree system is shown in Figure \ref{fig:local_method}-(a).
The master thread executes a $while(1)$ loop; In each iteration, it selects new nodes to send to worker threads, and checks for backup requests received from any worker in the worker-thread pool. The worker threads' processes are completely independent of one another; they only coordinate with the centralized master thread. The main advantage of the local-tree method is that it can overlap the computation of DNN inferences and in-tree operations by separating them into different hardware resources (Figure \ref{fig:local_method}-(b)); Also, for small-sized trees that can fit in last-level cache, the memory access latencies in in-tree operations are reduced compared to the shared-tree method. The disadvantage is that all the in-tree operations are completely serialized, leading to lower in-tree throughput.

\subsection{Adaptive Parallelism: System Overview}
The local-tree and shared-tree methods have tradeoffs that suit different scenarios. The intuition is that when DNN inference throughput is the bottleneck, the local-tree method should be favored to fully exploit the parallelism for independent Node Evaluations; when the number of workers becomes large or the tree is very deep such that the sequential in-tree operations become the bottleneck, the shared-tree method should be utilized to parallelize the in-tree operations between workers. In this work, we are motivated to take the best of both works and develop a tree-parallel DNN-MCTS implementation that is able to adaptively switch between the two methods. This implementation is facilitated with an empirical model to determine which method is best suited at compile time given an arbitrary DNN-MCTS algorithm specification and multi-core CPU device specification (later discussed in Section \ref{sec:perfmodel}).

To support adaptive parallelism that enables switching between the local-tree and shared-tree methods, we implement the DNN-MCTS program as shown in Algorithm 1. The program is an iterative process of data collection (Algorithm 1, lines 3-12) and DNN training (Algorithm 1, lines 13-15). Based on an input flag passed to the main program (Algorithm 1, lines 6-9), it selects between the shared-tree and local-tree methods, shown in Algorithm 2 and 3, respectively.

\begin{algorithm}

\DontPrintSemicolon
\caption{shared-tree based search}
\SetKwFunction{FMain}{get\_action\_prior\_s}
  \SetKwProg{Fn}{Function}{:}{}
  \Fn{\FMain{$environment$}}{
        $game \gets$ copy$(environment)$\;
        \For{\_\_ in num\_playouts} {
            add \texttt{threadsafe\_rollout}$(game)$ to thread pool\;
        }
        wait for threads to finish all work\;
        $action\_prior \gets$ normalized root's children list wrt visit count\;
        \KwRet\ $action\_prior$\;
  }
\SetKwFunction{GMain}{threadsafe\_rollout}
\SetKwProg{Pn}{Function}{:}{}
  \Pn{\GMain{$game$}}{
        $node \gets root$\;
        \While{$node$ is not leaf} {
            $node \gets node$'s child with highest UCT score\;
            $game$ execute the corresponding move\;
            obtain $lock$\;
            update $node$'s  UCT score with virtul loss\;
            release $lock$\;
        }
        $priors$, $value \gets $ \texttt{neural\_network\_simulate}$(game)$\;
        $node$ create children list according to $priors$\;
        obtain $lock$\;
        \texttt{backup}($node$, $value$)\;
        release $lock$\;
        \KwRet\;
  }
\label{alg:de}
\end{algorithm}

In the shared-tree method, a pool of threads is spawned to execute all the in-tree operations and DNN inferences in parallel. When a function is added to the thread pool (Algorithm 2, line 4), the input of the function is sent to an available thread, and the function is executed on the same thread. In the case of the shared-tree method, the function executed by each thread is 
``threadsafe\_rollout". It first traverses the tree from root to leaf, performing node selection, then performing node evaluation through ``neural\_network\_simulate", followed by node expansion and backup. During the virtual loss update and backup, multiple threads may share write accesses to the same nodes, so locks are used to ensure atomic accesses.

\begin{algorithm}
\DontPrintSemicolon
\caption{local-tree based search}
\SetKwFunction{FMain}{get\_action\_prior\_l}
  \SetKwProg{Fn}{Function}{:}{}
  \Fn{\FMain{$gomoku$}}{
        \texttt{rollout\_n\_times}($gomoku$, $num\_playouts$)\;
        $action\_prior \gets$ normalized root's children list wrt visit count\;
        \KwRet\ $action\_prior$\;
  }
\SetKwFunction{GMain}{rollout\_n\_times}
\SetKwProg{Pn}{Function}{:}{}
  \Pn{\GMain{$gomoku$, $num\_playouts$}}{
        \For{\_\_ in num\_playouts} {
            $node \gets root$\;
            \While{$node$ is not leaf} {
                $node \gets node$'s child with highest UCT score\;
                $game$ execute the corresponding move\;
            }
            add \texttt{neural\_network\_simulate}$(game)$ to thread pool\;
            \If{number of tasks in \text{thread pool} $\geq$ number of threads}{
                wait for a task to finish in the thread pool\;
                $priors$, $value \gets$ result of the task\;
                $node$ create children list according to $priors$\;
                \texttt{backup}($node$, $value$)\;
            }
        }
        \KwRet\;
  }
  \label{alg:c}
\end{algorithm}

In the local-tree method, a centralized master thread is responsible for all the in-tree operations, and a thread pool is spawned to execute all the DNN inferences asynchronously in parallel. Specifically, the master thread executes the ``rollout\_n\_times" (Algorithm 3, line 6-17). It repeatedly performs node selection, expansion, and backup, and assigns a ``neural\_network\_simulate" function as node evaluation request to the thread pool through a first-in-first-out queue. When all the threads are occupied by DNN inferences in the thread pool, the master thread waits until receiving a value for backup. Otherwise, it continues with the in-tree operation loop to generate node evaluation requests.

\subsection{Accelerator-offloaded DNN Inference}

Our implementation also supports offloading the DNN inferences onto a GPU. 
We utilize a dedicated accelerator queue for accumulating DNN inference task requests produced by the tree selection process. When the queue size reaches a predetermined threshold, all tasks are submitted together to the GPU for computation. Acceleration of DNN inferences is particularly important, especially when the total latency of in-tree operations is relatively small. However, it does require careful tuning of the communication batch size associated with the accelerator queue.

In the case of the shared-tree method, the communication batch size is always set to the number of threads employed (i.e., thread pool size). This is because the selection processes are parallel, resulting in the nearly simultaneous arrival of all inference tasks, leaving only a small gap to wait for the inference queue to be full.

The case of the local-tree method necessitates empirical tuning of the communication batch size. This is because the selection processes on the master thread are sequential and lead to long waiting times by the worker threads; submitting a small batch of inference tasks before the worker threads reach full capacity can help reduce accelerator waiting time, overlapping DNN inference computation with in-tree operations. Our empirical exploration of the communication batch size can be found in Section \ref{sec:dse_batch} and \ref{sec:dse_eval}.
\section{Performance Analysis for\\Adaptive Parallelism}
\label{sec:perfmodel}
\subsection{Performance Model}
In this section, we provide a theoretical analysis of the time performance to understand the tradeoff between the shared tree and local tree methods. The main parallel parameters that affect their performance include the number of threads, the latency of executing in-tree operations and inferences on each thread, and the data access and/or data transfer latencies.

Assuming the complete tree-based search process is conducted on a multi-core CPU with a thread pool size of $N$, the amortized latency for each iteration of the shared tree method on a multi-core CPU can be estimated as:
\begin{equation}
    T_{shared}^{CPU} \approx T_{\text{shared tree access}} \times N + T_{select} + T_{backup} + T_{DNN}^{CPU}
\end{equation}
The $T_{\text{shared tree access}}$ refers to the latencies that occurred in multiple threads accessing CPU-shared memory (DDR) as they traverse the same node. For selection and backup in a shared tree, this overhead is non-avoidable as all parallel workers start from the same root node. The in-tree operations latency and the DNN inference latency are summed up since they execute sequentially on each thread.

If we offload the batched DNN computations onto a GPU, the per-iteration latency can be estimated by replacing the DNN inference execution time with $T_{DNN}^{GPU}$, which contains the PCIe data transfer overhead and the actual computation time.
\begin{multline}
    T_{shared}^{CPU-GPU} \approx T_{\text{shared tree access}} \times N + T_{select} + T_{backup} \\+ T_{DNN}^{GPU}(batch=N)
\end{multline}

 The amortized latency for each iteration of the local tree method on a multi-core CPU can be estimated as:
\begin{equation}
    T_{local}^{CPU} \approx max((T_{select} + T_{backup}) \times N, T_{DNN}^{CPU})
\end{equation}
In the local tree method, the in-tree operations and DNN inferences are overlapped. Therefore, the per-iteration execution time is bounded by either the DNN inference latency or the total latency of the sequential in-tree operations. 
\begin{multline}
\label{eq:cent_cpu_gpu_perf}
    T_{local}^{CPU-GPU} \approx max\{(T_{select} + T_{backup}) \times N, \\T_{PCIe}, T_{DNN-compute}^{GPU}(batch=B)\}
\end{multline}

For batched DNN computations on GPU, we select a (sub-)batch size $B<N$ such that $\frac{N}{B}$ CUDA streams \cite{cudastream} are initiated, each CUDA stream bulk-processes the node evaluation (DNN inference) requests after $B$ loop counts of in-tree operations. Therefore, the timeline of the local Tree using a CPU-GPU platform can be visualized similarly to that depicted in Figure \ref{fig:cpugpu_lat}; The only differences are (1) the $N$ worker threads are replaced with $\frac{N}{B}$ CUDA streams, and (2) the blue-colored pipe communication arrows appear every $B$ iterations (instead of $1$ iteration) of in-tree operations.
\subsection{Design Configuration Workflow}
\label{sec:dse_batch}
To decide the parallel method and relevant design parameters (i.e., accelerator inference batch size) at compile time, we first obtain $T_{DNN}^{CPU}$, $T_{select}$ and $T_{backup}$ of a single worker on a single thread by profiling their amortized execution time on the target CPU for one iteration. 
The DNN for profiling is filled with random parameters and inputs of the same dimensions defined by the target algorithm and application. The $T_{select}$ and $T_{backup}$ are measured on a synthetic tree constructed for one episode (i.e., multiple iterations) with random-generated UCT scores, emulating the same fanout and depth limit defined by the DNN-MCTS algorithm. These design-time profiled latencies will provide a close prediction for the actual latencies at run time. 
We can also obtain $T_{DNN}^{CPU-GPU}$ including the computation and data migration latency. In our implementation, the tree is managed as a dynamically allocated array of node structs that resides in the CPU DDR memory. Therefore, we estimate $T_{\text{shared tree access}}$ as the DDR access latency documented for the target CPU device. These are plugged into the performance models for $T_{shared}^{CPU}$ and $T_{local}^{CPU}$ at compile time to decide the optimal parallel method for an arbitrary DNN-MCTS algorithm on a CPU.

For exploring the design space on a CPU-GPU platform, an additional parameter $B$ (i.e., number of cuda streams, each processing a sub-batch) can affect the performance of the local tree method. A naive method is to iterate over all the possible values for $B (B\in [1,N])$ and empirically run an episode to test the average latency of each iteration. However, this makes the design space exploration complexity linearly proportional to $N$ and hard to scale to very large multi-core and accelerator systems. To address this, we make the following observations to equation \ref{eq:cent_cpu_gpu_perf}:
\begin{itemize}
    \item $(T_{select} + T_{backup})$ remains constant or monotonically decreases with increasing $B$. This is because the Expand operation waits for a batch of inferences to complete the UCT score of the newly added nodes before they can be traversed in Backup and Selection. The higher the CUDA stream batch size $B$, the less frequently the nodes get available to be traversed (the frequency of making new node-UCT scores available is about once per $\frac{N}{B}$ loop counts on the Master Thread). This (increasing $B$) may in turn make the total tree depths traversed by Selection and Backup smaller due to less-frequent node insertions. Therefore, the first term of equation \ref{eq:cent_cpu_gpu_perf} should be a constant or monotonically decreasing sequence wrt $B\in\{1,...,N\}$.
    \item $T_{PCIe}$ is the time for transferring a total of $N$ data samples (i.e., DNN inference requests) between the CPU and GPU through a PCIe interconnection. It can be viewed as $\frac{N}{B}$ transfers, each transfer processes a batch of $B$ data samples.
    Each transfer is associated with a fixed communication and kernel launch latency $L$. Therefore, $T_{PCIe}$ can be modeled as $(\frac{N}{B})\times L + \frac{N}{\text{PCIe bandwidth}}$. Based on this model, $T_{PCIe}$ is expected to be a monotonically decreasing sequence wrt $B\in [1,N]$.
    \item $T_{DNN}^{GPU}(batch=B)$ is expected to monotonically increase with increasing $B$. This is because larger $B$ leads to higher computational workloads.
    
    \item Based on Equation \ref{eq:cent_cpu_gpu_perf}, the element-wise maximum of two monotonically decreasing sequences ($(T_{select} + T_{backup})$ and $T_{PCIe}$) is also a monotonically decreasing sequence.
    The element-wise maximum of this resulting monotonically decreasing sequence and a monotonically increasing sequence ($T_{DNN}^{GPU}(batch=B)$) should be a ``V-sequence" which is a sequence that first monotonically decreases, then monotonically increases wrt $B$.


\end{itemize}

Essentially, we want to search the design space of $B$ and find its value yielding the minimum execution time, i.e., $\argmin_B T_{local}^{CPU-GPU}$. Based on the above observations, this enables us to exploit the property of a ``V-sequence",
and develop an efficient algorithm to determine $B$ at design time. We achieve this by modeling the problem of finding the best-performing CUDA stream batch size $B$ as the problem of finding the minimum value of a ``V-sequence" $T$ ($T$ is the array of per-iteration latency across different values of $B\in\{1,...,N\}$). Instead of testing every possible value for $B\in [1,N]$, we can sample a subset with a reduced complexity of $O(\log N)$ as shown in Algorithm 4. Note that this is the mirroring problem of finding the maximum value of a bitonic sequence in $O(\log N)$ time using binary search \cite{williams1976modification}.

\begin{algorithm}
\DontPrintSemicolon
\caption{Exploring the optimal CUDA stream batch size $B$}
\SetKwFunction{GMain}{FindMin}
\SetKwProg{Pn}{Function}{:}{}
  \Pn{\GMain{$T$, $lo$, $hi$}}{
    \If{$lo == hi$ }{
        \KwRet $B\leftarrow lo$\; 
    }
    $mid=\frac{lo+hi}{2}$\;
    Test Run with $B=mid$ and $B=mid+1$\;
    Record amortized latency $T[mid], T[mid+1]$  \;   
    \If{$T[mid] \geq T[mid+1]$}{
        \KwRet \texttt{FindMin}($T$, $mid+1$,$hi$)\;
    }
    \Else{
        \KwRet \texttt{FindMin}($T$, $lo$,$mid$)\;
    }
  }
\label{alg:dse}
\end{algorithm}
Note that for each Test Run (Algorithm 4 line 5), we do not need to run the DNN-MCTS until policy convergence; we only profile the latency performance in a single move (i.e., get\_action\_prior functions in Algorithm 2 and 3). This is because each move made in the complete DNN-MCTS training loop has the same amount of computations.
\section{Evaluation}
\subsection{Experiment Setup}
\textbf{Benchmark and hyper-parameters: }
We use the Gomoku game benchmark \cite{yan2018hybrid} to evaluate the performance of our proposed method. The board size (i.e., size of the input state to the policy/value network)  is 15$\times$15, the neural network is composed of 5 convolution layers and 3 fully-connected layers; The tree size limit per move is 1600 (i.e., The total number of selection-expansion-inference-backup operations performed per agent-move is 1600).

\textbf{Hardware platform specifications: } We use the AMD Ryzen Threadripper 3990X @ 2.2GHz as our target CPU platform. It has 64 cores (2 threads per core). The last-level cache size is 256 MB, and has a $8\times 32$-GB DDR4. The CPU is connected with a NVIDIA RTX A6000 GPU through PCIe 4.0.

\textbf{Evaluation metrics: } We conduct experiments to evaluate both the speed and parallel algorithm performance. The speed is measured through (1) the amortized per-worker-iteration latency in the tree-based search stage (Section \ref{sec:evallaat}), obtained by running and averaging all the 1600 iterations for making a move; and (2) the overall training throughput (Section \ref{sec:evalthrpt}) in terms of processed samples/second, obtained by $\frac{\text{Number of samples processed per episode}}{\sum(\text{Tree-based search time + DNN update time})}$. Note that one sample is obtained by executing all 1600 rounds of in-tree operations and DNN inferences in a move. The algorithm performance (Section \ref{sec:evalalg}) is measured using the loss of the DNN (Equation \ref{eq:loss}). The lower the loss, the more accurately the DNN is able to predict the probability of winning at each state and action, and the better the MCTS at guiding the moves toward the winning state. 

\subsection{Design Exploration of Host-Accelerator Communication Batch Size}
\label{sec:dse_eval}
\begin{figure}[h]
    \centering
    \includegraphics[width=6.5cm]{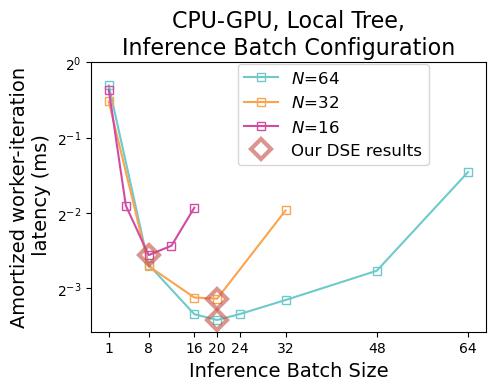}
    \caption{Design Exploration of Inference Batch Size}
    \label{fig:dse_exp}
\end{figure}

We show the performance obtained during the design configuration process for choosing the CUDA stream batch size $B$ in Figure \ref{fig:dse_exp}, specific to the local-tree method mapped to a CPU-GPU heterogeneous platform. We only perform this design exploration for the cases when the available number of workers $N\geq 16$. This is because $N\geq 16$ is the threshold where the shared-tree method starts to outperform the local-tree method with full-batched (batch size$=N$) inferences on GPU (later discussed in Section, Figure \ref{fig:cpugpu_lat}), and the question of whether choosing an alternative batch size could help improve the local-tree performance arises. 
We can observe that at smaller batch sizes, sub-batches of inferences are serialized, which hinders the performance. The extreme case is at batch size$=1$, where 
the serial inferences dominate the runtime, making the amortized iteration latency high such that even changing $N$ does not affect the performance. 
At larger batch sizes, inferences are parallelized with a higher degree on the GPU, but the inference request is made after waiting for all the serial in-tree operations to complete on the master thread, leading to a large overhead. The extreme case is at batch size$=N$, the GPU waits for all the $N$ before it can start the computation; the $N$ in-tree operations at the master thread is a non-trivial overhead such that they contribute to higher amortized latency at $N=64$ compared to $N=16$ or $32$. Our design exploration finds the balance point where there are enough inferences within each sub-batch to saturate GPU parallelism, while enough requests are also made across sub-batches such that the GPU computation can overlap with the computations on the CPU master thread (i.e., GPU does not have to be idling and waiting for CPU computation to finish). Based on our test runs, the optimal batch sizes are $8$ when $N=16$, and $20$ when $N=32$ or $64$. 

\subsection{Tree-based Search Iteration Latency}
\label{sec:evallaat}
\begin{figure}[h]
    \centering
    \includegraphics[width=6.5cm]{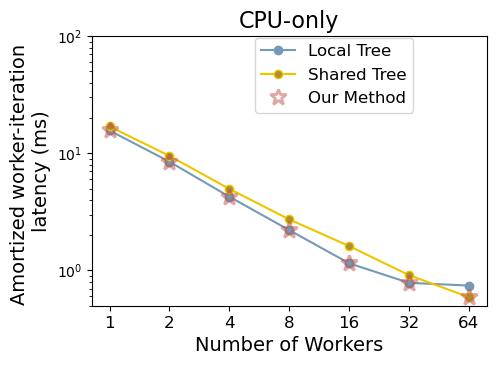}
    \caption{Iteration latency, CPU-only}
    \label{fig:cpu_lat}
\end{figure}

\begin{figure}[h]
    \centering
    \includegraphics[width=6.5cm]{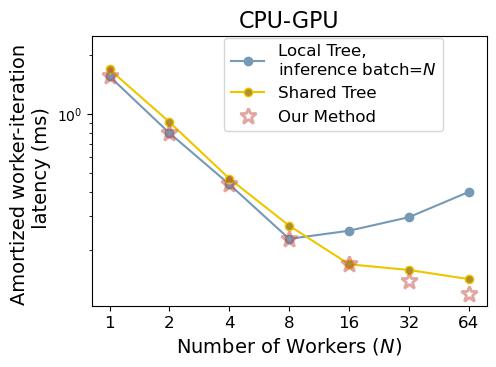}
    \caption{Iteration latency, CPU-GPU, batched inference}
    \label{fig:cpugpu_lat}
\end{figure}

We plot the amortized per-worker-iteration latency in the tree-based search stage in Figure \ref{fig:cpu_lat} and \ref{fig:cpugpu_lat}. Note that a worker iteration is one round of Node Selection, Node Expansion, Node Evaluation (DNN inference), and BackUp executed by one worker. In each move, 1600 such worker-iteration are executed by all the $N$ parallel workers. We obtain the amortized per-worker-iteration latency by dividing the total time for a move by 1600. The higher $N$ is, due to more parallelism exploited, the lower the total time for a move (and the amortized per-worker-iteration latency) is.
For the CPU-only implementation, each worker is assigned a separate CPU thread for performing one node evaluation (i.e., DNN inference). In Figure \ref{fig:cpu_lat}, we observe that under different configurations (number of workers used), the optimal method can be different. Our method using adaptive parallelism is able to always choose the optimal method, achieving up to $1.5\times$ speedup compared to either the local tree or the shared tree baselines on the CPU-only platform.
For the CPU-GPU implementation, a communication buffer is used to collect a batch of node evaluation requests before sending them to the GPU for performing a batched DNN inference. In Figure \ref{fig:cpugpu_lat}, we observe that if we set the buffer (batch) size, the amortized latency using the local tree method gets higher as $N$ increases over 16. At $N=16$, our implementation chooses the shared tree method. At $N=32$ and 64, using the optimal batch size returned by Algorithm 4, the local tree method combined with overlapped GPU inferences outperforms the shared tree method with full-batched GPU inferences. Overall, on a CPU-GPU heterogeneous platform, our method using the adaptive parallelism achieves up to $3.07\times$ speedup compared to either the local tree or the shared tree baselines.

\subsection{Throughput Analysis}
\label{sec:evalthrpt}
\begin{figure}[h]
    \centering
    \includegraphics[width=7.5cm]{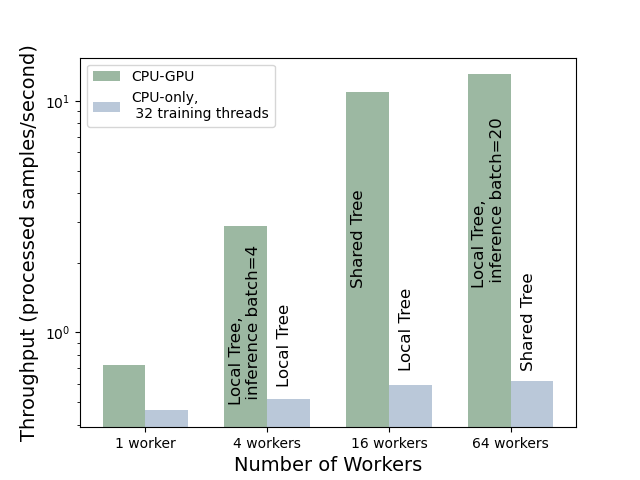}
    \caption{Training throughput under optimal configurations}
    \label{fig:throughput}
\end{figure}

We plot the overall DNN-MCTS training throughput (processed samples per second) for both the CPU-only and CPU-GPU platforms in Figure \ref{fig:throughput}, varying the number of workers used in the tree-based search. 
The throughput numbers are obtained by applying the optimal parallel method and design configuration returned by our design configuration workflow.
Overall, CPU-GPU implementations show 
higher throughput compared to CPU-only implementations.
In the CPU-GPU implementations, the tree-based search process produces samples and the training process (completely offloaded to GPU) consumes samples.
The training process execution time is hidden by the tree-based search time, especially when there is a small number of workers such that the in-tree operations and DNN inferences become the bottleneck. As the number of workers increases, we observe near-linear improvements in throughput, since the time spent producing the same number of samples for training is reduced. When the number of agents increases above 16, the tree-based search time is reduced to the extent that it is lower than the training time. As a result, the throughput improvement becomes less obvious.
In the CPU-only implementations, given the limited number of available CPU hardware threads, we are able to allocate 32 threads for conducting training on the CPU (these are different threads than those used for DNN-MCTS parallel workers). In contrast to GPU-accelerated training, CPU-based DNN training now becomes the bottleneck even for a small number of DNN-MCTS workers. With a different number of workers allocated to the tree-based search process, the compute power provided to the training process is fixed (32 threads). Therefore, the throughput improvements from increasing the number of DNN-MCTS workers are not as scalable as the CPU-GPU implementations. 
Still, we are able to adaptively choose the best-performing parallel method and design configurations. The optimal methods used at different hardware platforms and available resources (i.e., number of workers) are annotated in Figure \ref{fig:throughput}.

\subsection{Algorithm Performance}
\label{sec:evalalg}

\begin{figure}[h]
    \centering
    \includegraphics[width=8cm]{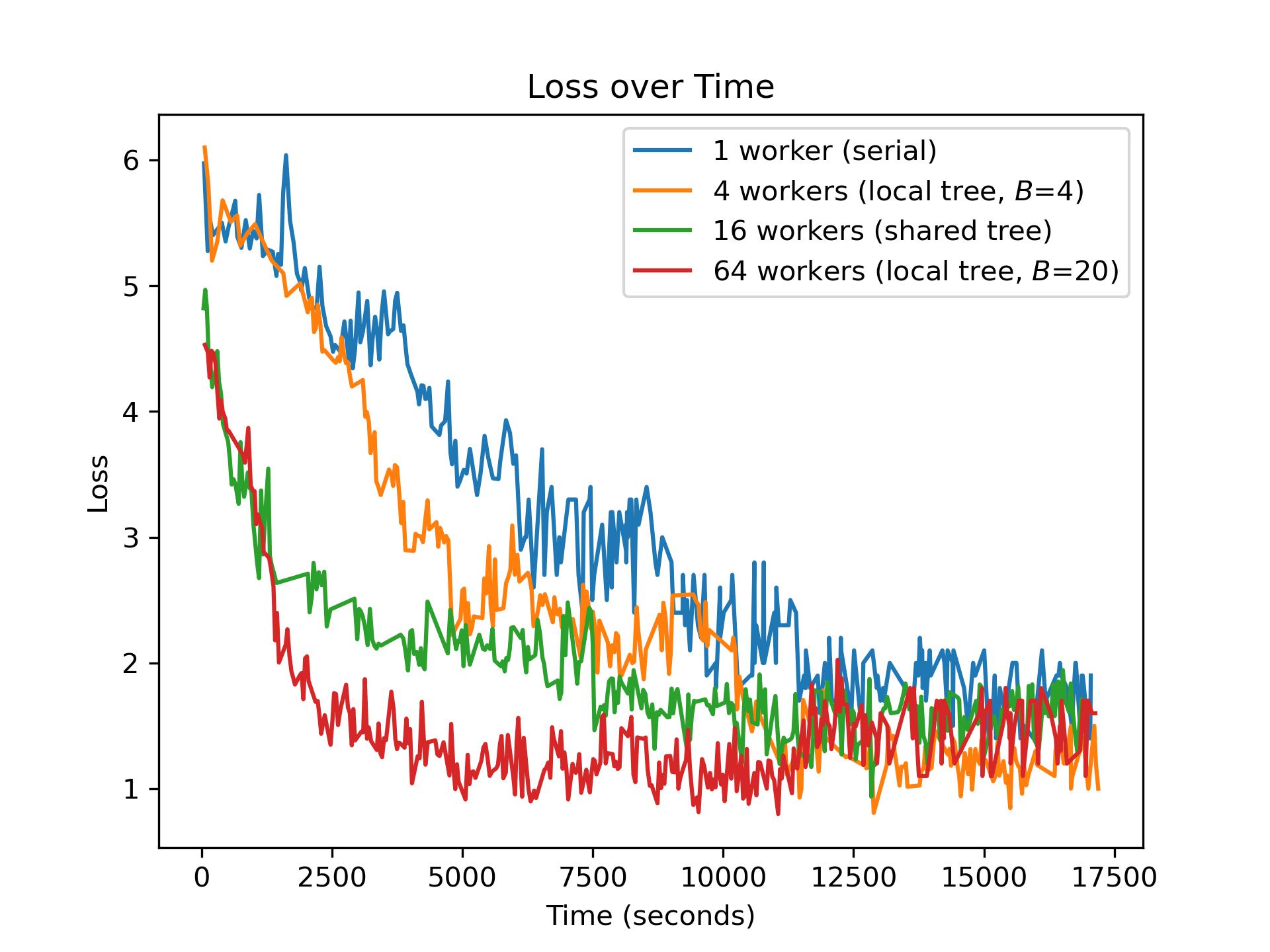}
    \caption{DNN loss over time, using the optimal parallel methods returned by our Design Configuration across different number of parallel workers}
    \label{fig:loss}
\end{figure}

We show the DNN loss over time as the measurement of parallel DNN-MCTS training algorithm performance in Figure \ref{fig:loss}. The experiments are conducted on the CPU-GPU platform using the optimal parallel configurations for 4, 16, and 64 workers. As we introduce parallel workers for the tree-based search, the algorithm is modified. This is because in the serial tree search, every iteration accesses the most up-to-date tree information modified by the previous iteration; while in the tree-parallel implementations, a worker traversing the tree may not obtain the newest node UCT scores because the node evaluation (i.e., DNN inference) of other workers have not completed. The more parallel workers are used, the higher the effect is from such obsolete-tree-information. As a result, the training samples generated (states traversed and actions taken based on tree search) in the parallel version are not the same as the 1-worker serial baseline. Still, the converged loss is not negatively impacted by increasing parallelism, as shown in Figure \ref{fig:loss}. Additionally, the convergence curve is steeper, meaning the time taken to reach the same converged loss is reduced using the optimal parallel configurations of our adaptive parallel implementations.








\section{Conclusion}
In this work, we proposed a novel implementation for DNN-MCTS that adaptively chooses the optimal parallel scheme for the MCTS component on the CPU. We also analyzed the performance on a CPU-GPU platform and proposed an efficient method to search for the optimal communication batch size interfacing the MCTS component and DNN operations. 
By experimenting on a CPU-only and CPU-GPU platform using a Gomoku game benchmark, we observed up to $1.5 \times$ and  $3.07\times$ speedup using our adaptive parallelism compared to existing fixed-parallelism methods.
Our method and performance models are general and can also be adopted in the context of many other types of accelerators for DNN inference and training ( FPGAs, ASICS (e.g., TPUs), etc.) in the future.

\newpage
\bibliographystyle{ACM-Reference-Format}
\bibliography{main}

\end{document}